\documentclass[runningheads]{llncs}
\usepackage[T1]{fontenc}
\usepackage{graphicx}
\usepackage{svg}
\usepackage{amsmath}
\usepackage{amssymb}
\usepackage{comment}
\usepackage{mathtools}
\usepackage{multirow}
\usepackage{booktabs}
\usepackage{hyperref}
\usepackage{subcaption}
\footnotesize
\tiny
\usepackage{graphicx}
\begin{document}

\title{Enhanced Diagnostic Fidelity in Pathology Whole Slide Image Compression via Deep Learning}
\titlerunning{Perceptual Image Quality for lossy compressed WSI}
%
%
\author{Maximilian Fischer\inst{1,2,3} \and
Peter Neher\inst{1,2,11} \and
Peter Sch\"uffler\inst{6,7} \and
Shuhan Xiao\inst{1,4} \and
Silvia Dias Almeida\inst{1,3} \and
Constantin Ulrich\inst{1,3,12} \and
Alexander Muckenhuber\inst{6} \and
Rickmer Braren\inst{8} \and
Michael G\"otz\inst{1,5} \and
Jens Kleesiek\inst{9,10} \and
Marco Nolden\inst{1,11} \and
Klaus Maier-Hein\inst{1,3,4,11,12} 
}
\authorrunning{Fischer et al.}

%
\institute{
Division of Medical Image Computing, German Cancer Research Center (DKFZ), Heidelberg, Germany\and  
German Cancer Consortium (DKTK), partner site Heidelberg\and               
Medical Faculty, Heidelberg University, Heidelberg, Germany\and            
Faculty of Mathematics and Computer Science, Heidelberg University, Heidelberg, Germany\and
Clinic of Diagnostics and Interventional Radiology, Section Experimental Radiology, Ulm University Medical Centre, Ulm, Germany\and 
TUM School of Medicine and Health, Institute of Pathology, Technical University of Munich, Munich, Germany\and 
TUM School of Computation, Information and Technology, Technical University of Munich, Munich, Germany\and
Department of Diagnostic and Interventional Radiology, Faculty of Medicine, Technical University of Munich, Munich, Germany \and
Institute for AI in Medicine (IKIM), University Medicine Essen, Essen, Germany\and 
German Cancer Consortium (DKTK), partner site Essen \and                   
Pattern Analysis and Learning Group, Department of Radiation Oncology, Heidelberg University Hospital, Heidelberg, Germany\and      
National Center for Tumor Diseases (NCT), NCT Heidelberg, a partnership between DKFZ and University Medical Center Heidelberg\\     
\email{maximilian.fischer@dkfz-heidelberg.de \\}}

\maketitle              
This is a modified preprint
Find the original here \\
\doi{10.1007/978-3-031-45676-3_43}
\begin{abstract}
Accurate diagnosis of disease often depends on the exhaustive examination of Whole Slide Images (WSI) at microscopic resolution. Efficient handling of these data-intensive images requires lossy compression techniques. This paper investigates the limitations of the widely-used JPEG algorithm, the current clinical standard, and reveals severe image artifacts impacting diagnostic fidelity. \\
To overcome these challenges, we introduce a novel deep-learning (DL)-based compression method tailored for pathology images. By enforcing feature similarity of deep features between the original and compressed images, our approach achieves superior Peak Signal-to-Noise Ratio (PSNR), Multi-Scale Structural Similarity Index (MS-SSIM), and Learned Perceptual Image Patch Similarity (LPIPS) scores compared to JPEG-XL, Webp, and other DL compression methods. \\

\end{abstract}
\section{Introduction}

Spatial formations and overall morphological characteristics of cells serve as fundamental elements for accurately diagnosing various diseases. The detection of global cellular patterns requires a comprehensive examination of tissue samples, posing a significant challenge due to the potentially very large size of these specimens. To address this challenge, Whole Slide Imaging (WSI) techniques provide high-resolution images that capture cells at a microscopic level. However, efficient handling, storage, and transmission of these files necessitate the use of lossy image compression algorithms.

Among the lossy compression algorithms employed by WSI vendors, JPEG80 is the most common one. But even with initial compression, the resulting file sizes remain a major obstacle in effectively transmitting or storing WSI. Thus the further compression of WSI is of significant interest and a field of active research. The fundamental burden is to compress images with high image quality, as visual inspection by pathologists continues to be the clinical standard, and accurate diagnoses heavily rely on the ability to examine the WSI with exceptional visual quality. Despite its critical importance, previous studies have primarily focused on evaluating the impact of compression on deep learning (DL) models \cite{chen_quantitative_2020,ghazvinian_zanjani_impact_2019} instead. 

In this paper, we present a novel DL-based compression scheme specifically tailored for histopathological data and compare it for several metrics to quantify perceptual image quality for pathological images. The proposed method ensures high levels of image quality, even at high compression ratios, by enforcing similarity between deep features from the original and the reconstructed image that are extracted by a contrastive pre-trained DL model. We show that our method outperforms traditional codecs like JPEG-XL or WebP, as well as other learned compression schemes like \cite{bmshj2018-factorized} in terms of image quality, while we also achieve higher compression ratios. \\
We compare our method to other approaches using various state-of-the-art image quality metrics and suggest establishing the Learned Perceptual Image Patch Similarity (LPIPS) \cite{LPIPS} metric from the natural scene image domain also for the pathology domain as image quality measure.\\ 
The presented findings have the potential to significantly impact the field of pathological image analysis by enabling higher compression ratios while preserving image quality. We show that our method yields much higher perceptual image quality metrics for the same compression ratio than the compared compression schemes.

\section{Methods}

\subsection{Learned Lossy Compression} 

Lossy image compression methods are usually modeled as follows: An image is projected into an alternative (latent) representation, where pixel values are quantized to reduce the storage requirements for each pixel, accepting distortions in the image due to information loss at the same time. 
For a learned lossy image compression scheme, the required latent representation is usually determined by an autoencoder architecture \cite{TheisPioneering}. In this setting, an encoder model $E$ generates quantized latents $y=E(x)$, from which the decoder model $D$ decodes a lossy reconstruction $x'$. The quantized latents are stored via a shared probability model $P$ between the encoder and decoder part that assigns a number of bits to representations according to their occurrence frequency in the dataset, which is determined by the discrete probability distribution obtained by $P$. The probability model together with an entropy coding scheme enables lossless storing of $y$ at a bitrate $r(y)= -log (P(y))$, which is defined by Shannon's cross entropy \cite{Shannon}. The distortion $d(x,x')$ measures the introduced error between the original image and the reconstruction, which can be determined by metrics such as the Mean Squared Error $d=\mathit{MSE}$. Learned codecs usually parametrize $E$, $D$ and $P$ as convolutional neural networks (CNNs) and the problem of learned image compression can be summarized as bitlength-distortion optimization problem~\cite{TheisPioneering}.

\subsection{Image quality metrics} 
Lossy image compression schemes universally result in a degradation of image quality as compression factors increase. To assess the performance of compression algorithms, image quality metrics serve as numerical measures. However, a prevailing issue is that no existing metric captures all kinds of image distortions.

For instance, when subjected to high compression factors, JPEG compression exposes blocking artifacts that the Peak Signal-to-Noise Ratio (PSNR) fails to detect effectively. Conversely, learned image quality metrics exhibit reduced sensitivity to color distortions, which highlights the challenges in achieving a comprehensive assessment of image compression performance \cite{Hific}. Thus compression schemes always should be evaluated by multiple image quality metrics. 

\subsubsection{Peak-Signal-to-Noise Ratio} 
The most common metric is the Peak-Signal-to-Noise Ratio (PSNR), which is derived by the mean squared error (MSE). The MSE compares the "true" pixel values of the original image versus the reconstructed pixel values of the decompressed image. By computing the mean of the squared differences between the original image and the noisy reconstruction, the difference between the reconstruction and the original image can be measured. Assuming a noise-free original image $x$ with the dimensions $m \times n$ and a noisy approximation $x'$ after decoding the quantized latents, the MSE and PSNR are defined as follows: 
\begin{equation}
\label{eqn:LossFunction}
\mathit{MSE} = \frac{1}{m n} \sum_{i=0}^{m-1} \sum_{j=0}^{n-1} \lbrack x(i,j) - x'(i,j) \rbrack^{2}\, \text{and}\, \mathit{PSNR} = 10\cdot log_{10} (\frac{\mathit{MAX}_{x}^{2}}{\mathit{MSE}})
\end{equation}

with $\mathit{MAX_x}$ being the largest possible value that can occur in the original image $x$. 

\subsubsection{Multi Scale Structural Similarity Index Measure}

It has been shown that assuming pixel-wise independence in structured datasets like images leads to blurry reconstructions \cite{Hific}. In particular, the MSE is not able to capture connections between neighboring pixels. In contrast, the Multi Scale Structural Similarity Index Measure (MS-SSIM) \cite{MSSSIM} takes into account that pixels have strong dependencies among each other, especially when they are spatially close. The MS-SSIM is based on the $\mathit{SSIM}$ metric which is calculated across multiple windows and scales between two images. For two windows $u$ and $u'$, which are located in $x$ and $x'$ respectively, the MS-SSIM is calculated as \cite{MSSSIM}:
\begin{equation}
\label{eqn:LossFunction}
\mathit{SSIM} (u,u') = \frac{(2\mu_{u}\mu_{u'} + c_{1}) (2\sigma_{u,u'} + c_{2})}{(\mu_{u}^{2} + \mu_{u'}^{2} + c_{1})(\sigma_{u}^{2} + \sigma_{u'}^{2} + c_{2})} ,
\end{equation}

with the averaged intensities in the given block window $\mu$, the variance $\sigma$, the covariance $\sigma_{u,u'}$ and the parameters $c_1,c_2$ for stabilizing the division by $c_1=(k_1L)^2$ and $c_2=(k_2L)^2$ where we set $k_1=0.01,k_2=0.03$ and $L$ models the dynamic range of the pixel values. More details can be found in \cite{MSSSIM}.

\subsubsection{Learned Image similarity}
In contrast to traditional metrics like MSE and MS-SSIM, Learned Perceptual Image Patch Similarity (LPIPS) is a perceptual metric that has been trained to align more closely with human interpretation of image quality \cite{LPIPS}. Unlike the aforementioned metrics, LPIPS does not measure the distance between two images solely at the pixel level. Instead, it quantifies the similarity of two images based on their embeddings that are obtained from multiple depths of a pre-trained model with depth $L$. In \cite{LPIPS} it is suggested to use an ImageNet \cite{russakovsky_imagenet_2015} pre-trained model to generate latent embeddings for the original image $x$ and the distorted image $x'$. The LPIPS distance is quantified by the averaged $l_{2}$ distance between multiple depths $l$ which are aggregated by learned weighting coefficients $\{\alpha_{l}\}_{l}^{L}$. For the distance, the embeddings are unit-normalized in the channel dimensions and the metric is defined as: 
\begin{equation}
\label{eqn:LPIPS}
d_{\text{LPIPS}} (x,x')^{2} = \sum_{l=1}^{L} \alpha_{l} \vert\vert \phi_{l}(x) - \phi_{l}(x')\vert\vert_{2}^{2},
\end{equation}
where $\phi_{l}$ is the feature map at layer $l$.

\subsection{Deep perceptual guidance}

Inspired by approaches like \cite{LPIPS}, we incorporate a learned feature similarity in the loss function during training of our lossy compression autoencoder. In contrast to previous work, we use a contrastive pre-trained model from the pathology domain to compute distances between embeddings. We hypothesize that the infused domain knowledge of pathological images leads to more realistic reconstructions of histopathological images. Furthermore, we use a more lightweight implementation than LPIPS for calculating distances between embeddings to reduce the added computational complexity. In particular, we use the contrastive pre-trained model from \cite{CIGA2022100198} and compute $\ell_{2}(C(x),C(x'))$, which is the $\ell_{2}$ distance between the feature vectors from $x$ and $x'$, both extracted with the model $C$ from \cite{CIGA2022100198}. The weighted $\ell_{2}$ distance measure, together with the baseline MSE distance and the targeted bitrate yields the final loss function: 
\begin{equation}
\label{eqn:LossFunction}
\mathcal{L}_{EG} = r+\lambda\cdot d = \mathbb{E}_{x \sim p(x)} \lbrack \underbrace{ \lambda r(y)}_\text{bitrate} + \underbrace{d(x,x') + \psi \ell_{2}(C(x),C(x'))}_\text{image similarity}\rbrack
\end{equation}

\section{Experiments}\label{Experiments}
\subsection{Dataset}\label{Dataset}
To determine the performance of a lossy compression autoencoder in the histopathology domain, we collect a diverse dataset containing various tissue types \cite{BreaKHis,LungColonData,KatherDataSet}. In table \ref{Dataset}, an overview of the collected dataset for this study is presented. In this paper, we investigate the impact of further compression schemes, beyond the initial compression during image acquisition. Thus all images were initially JPEG80 compressed, without any further compression besides that. We split the dataset on a slide-level label into training and test set.
To compare the performance of different compression schemes, we generate compressed versions of the test set at various bit rates. For each bit rate, we evaluate the image quality metrics between the original image and the compressed version. As traditional codecs we implement JPEG, WebP and JPEG-XL. 
\begin{table}[]
\centering
\caption{Details of the datasets used for this work.}
\label{table:Dataset}
\begin{tabular}{cccc}
Dataset      & BreaKHis & Colon1 & Colon2 \\ \hline
Source       & \cite{BreaKHis}         & \cite{LungColonData}       &  \cite{KatherDataSet}       \\
Tissue & Breast & Colon & Colon \\
Images       &   1639         &500        &10        \\
SampleSize   &700x460          &768x768        &5000x5000        \\
Tile Size    &224x224          &224x224        &224x224        \\
Tiles        &10158          &10005        &4840        \\
\\
\begin{tabular}{l}

\end{tabular} &   \includegraphics[width=20mm,height=20mm]{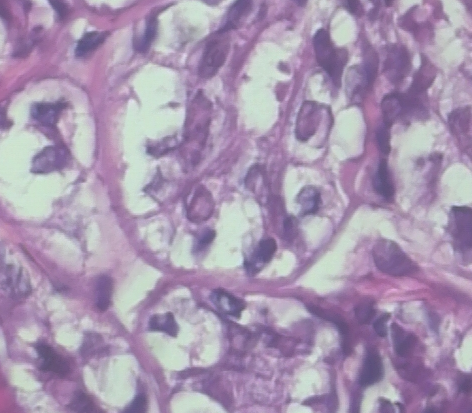}       &\includegraphics[width=20mm,height=20mm]{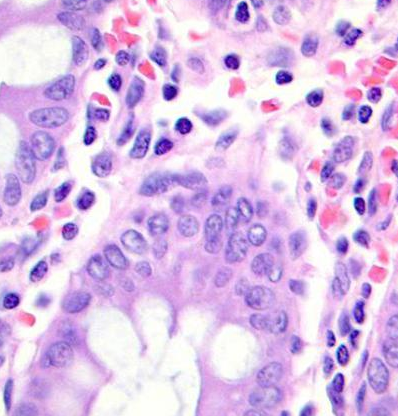}        &       \includegraphics[width=20mm,height=20mm]{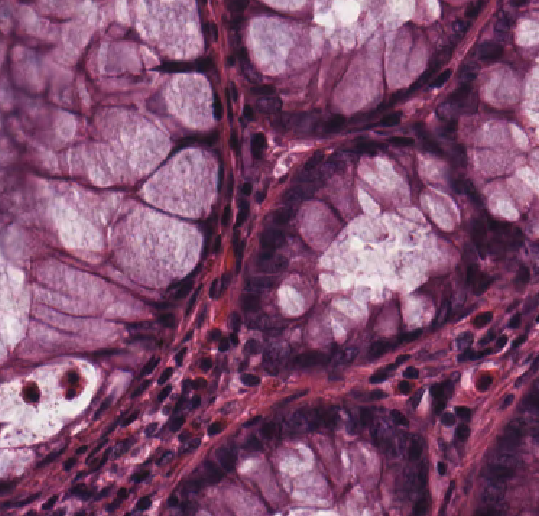}\\
\end{tabular}
\end{table}

\subsection{Neural Image Compression Models}

As a compression autoencoder, we implement the model "bmshj-factorized" from \cite{bmshj2018-factorized}. During training, we use the implementation from equation \ref{eqn:LossFunction} for training our model to generate realistic reconstructions. We used \cite{begaint2020compressai} as a framework to implement our method. For each targeted bit rate, we train a separate model, by sampling values for $\lambda$ from $\lambda=[0.001,...,0.1]$. Although it might be technically feasible to train one model for various bit rates \cite{Multiplebpp1,Multiplebpp2,Multiplebpp3}, this enables us to target the desired bit rates more accurately. The models are trained with an initial setting of $\psi=0.5$ for 90 epochs, which is increased to $\psi=0.7$ for another 60 epochs. In total, we train all models for approximately 4M steps using the Adam optimizer and we decrease the initial learning rate of 0.0001 whenever equation \ref{eqn:LossFunction} yields no further improvement. We use a tile size of $224 \times 224$ pixels and a batch size of 4. During training, we apply random horizontal and vertical flips, as well as color jittering. To approximate quantization during testing, we employ random noise during training like suggested in \cite{TheisPioneering}. During testing, we evaluate the performance of the model with the PSNR, as well as the perceptual image quality metric MS-SSIM and LPIPS. We refer to this training scheme as Supervised Pathologic L$_{2}$ Embeddings ($SPL_{2}E$). To evaluate the impact of the deep pathologic feature supervision, we train the same model also without the additional feature distance $\psi$, which we refer to as \textit{Baseline}. Furthermore, to evaluate the impact of training on pathology data in general for compression models, we also compare the performance of the same model that is trained on the Vimeo dataset \cite{Vimeo}. This dataset contains a large collection of natural scene images. We refer to this model as \textit{Vimeo}.

\section{Results} \label{Results}

\subsection{Comparison of Image Quality Metrics}
\begin{figure}[h!]
    \centering
    \includegraphics[width=\textwidth]{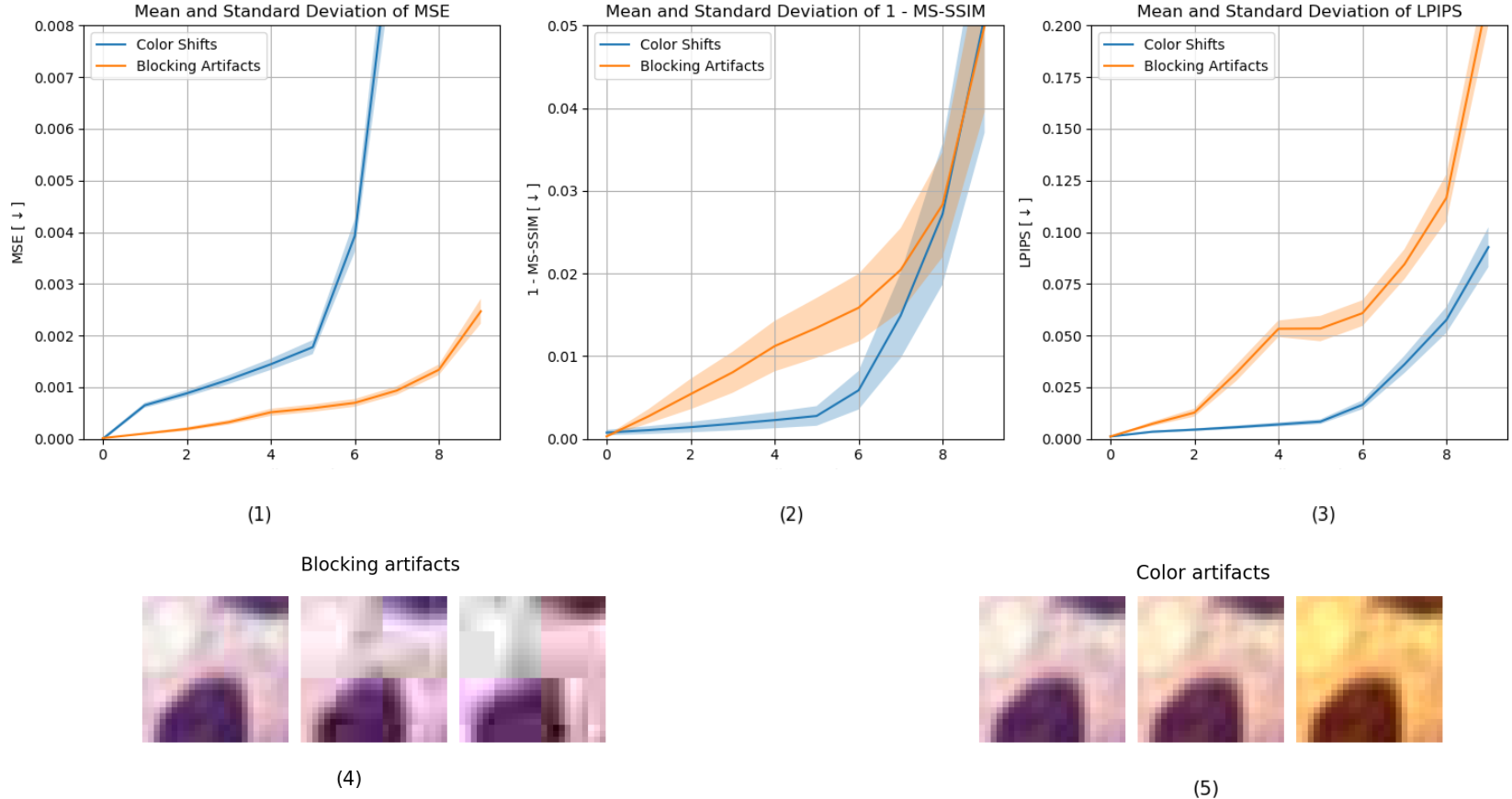}
    \caption{Visualization of the three metrics for color and blocking artifacts as image distortions. We report $1-$MS-SSIM for better visual inspection (1-3). Below the plots, the levels of distortions are visualized for the blocking (4) and the color (5) artifacts. For visualization purposes, we show the quality factors $90$, $40$ and $5$ for the blocking artifacts. }
    \label{Figure:MetricsProgression}
\end{figure}

Figure \ref{Figure:MetricsProgression} illustrates the metric scores for two types of potential distortions encountered in lossy image compression: color shifts and blocking artifacts. The x-axis represents the degree of distortion, which corresponds to the quantization level or quality factor for lossy JPEG compression for blocking artifacts, and a bias between $0$ and $50$ added in the Lab color space for color shifts. For the blocking artifacts, we sample values between $90$ and $10$ as quality factors for the JPEG compression. The respective metric is then computed between the distorted image and the original image. The presented results are the average scores obtained from evaluating 10 sample images from our test set.
The figure shows that the Mean Squared Error (MSE) metric is highly sensitive to color distortions, whereas the Learned Perceptual Image Patch Similarity (LPIPS) metric effectively detects and quantifies blocking artifacts.


\subsection{Quantitative PSNR, MS-SSIM and LPIPS metrics} \label{numericRes}
\begin{figure}[h!]
    \centering
    \includegraphics[width=\textwidth, trim=10pt 0 10pt 0, clip]{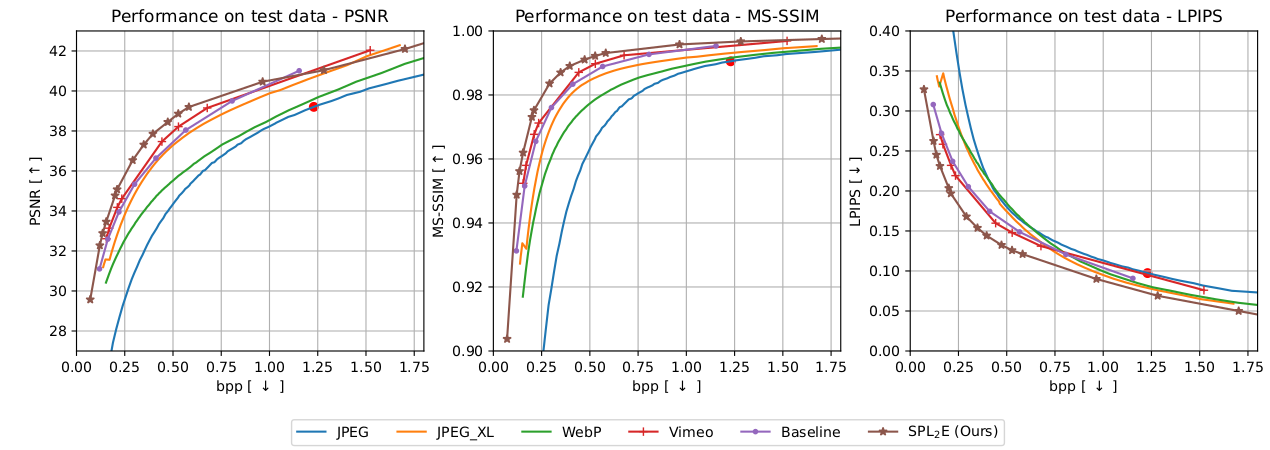}
    \caption{Rate-distortion and -perception curves on the test set for various metrics. Arrows ($\uparrow$), ($\downarrow$) indicate whether high or low scores are preferable. The level of compression is determined by the bits-per-pixel (bpp) value.}
    \label{Figure:MetricComp}
\end{figure}

Figure \ref{Figure:MetricComp} shows the results obtained by the compression schemes for the resulting bit rate and perceptual image quality metrics. Arrows in the figure indicate whether a high or low value of the metric is better. The resulting bitrate the compression with JPEG80 would reach is marked as a red dot in the plots. Our results show that our proposed method performs equal or better than the compared methods, while JPEG and WebP are showing the worst results across all metrics. Especially between compression ratios of $0.2\,bpp$ and $1.1\,bpp$, the proposed methods outperforms all compared approaches. We show that our model is able to achieve 76\% smaller file sizes than JPEG, while maintaining the same amount of image quality (compare $0.4\,bpp\,$vs.$\,1.2\,bpp$ for MS-SSIM $=0.99$).

\subsection{Qualitative Assessment}
\begin{figure}[h!]
    \centering
    \includegraphics[width=\textwidth, trim=0cm 0cm 0cm 0cm, clip]{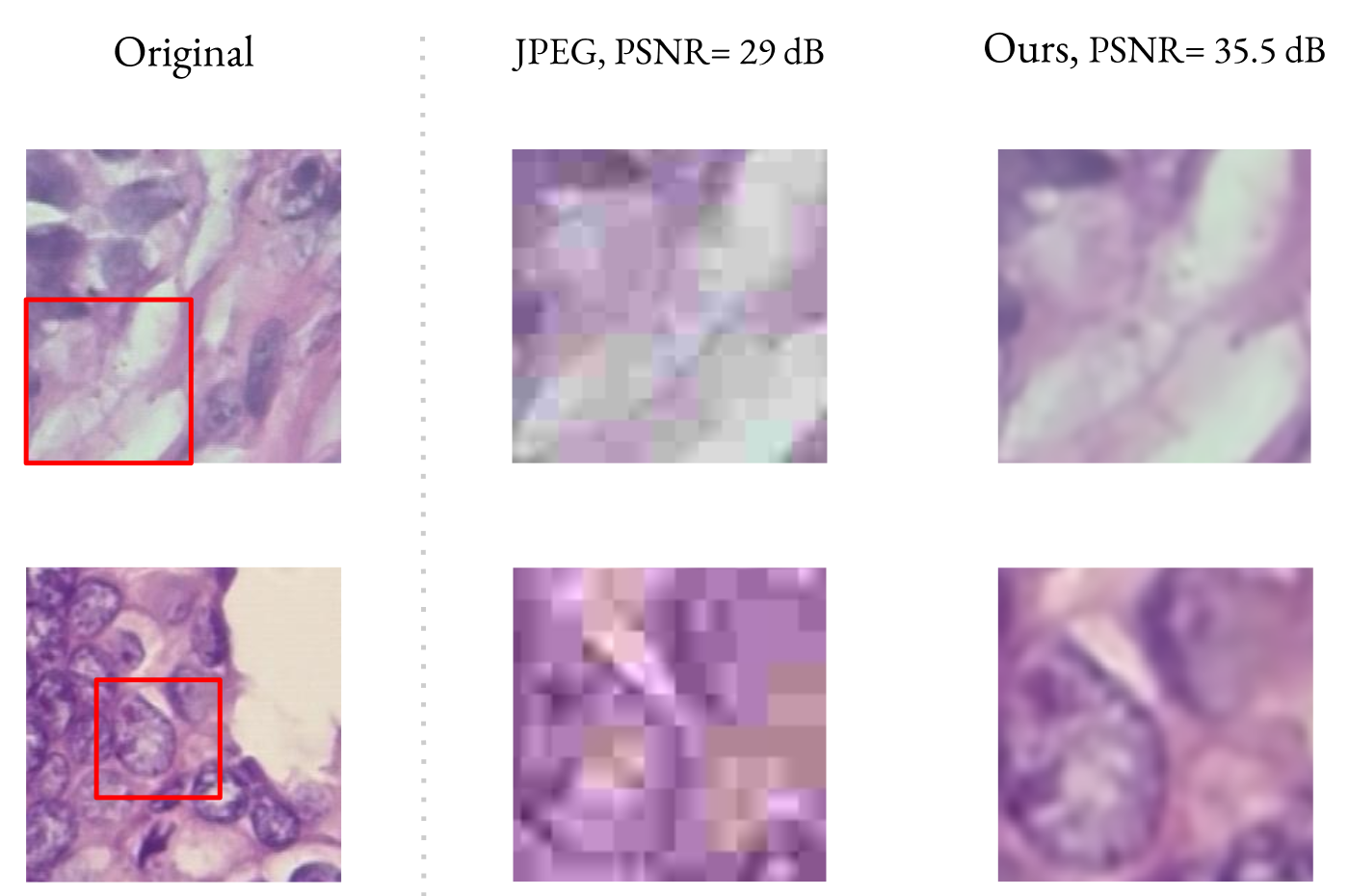}
    \caption{The proposed model achieves much higher perceptual image quality than JPEG for the same compression ratio (here 0.25 bpp). Left: Two different original patches. Middle: JPEG compression, Right: Ours. Patches are taken from \cite{BreaKHis,LungColonData}}
    \label{Figure:Qualitative}
\end{figure}

We show the qualitative reconstruction result of two exemplary image patches in figure \ref{Figure:Qualitative}, at the same compression ratio. The figure shows the reconstructed result of the proposed method. Further images showing reconstructed images of the other compression schemes can be found in figure 1 in the supplementary material. The figure shows that the proposed method achieves much higher perceptual image quality at the same resulting bit rate than JPEG. The figure demonstrates that our approach exhibits no color smearing (indicated by the white background in the left images) and maintains clear visibility of cellular structures (right images).

\section{Discussion}
This paper shows how a DL-based lossy compression scheme is fine-tuned on pathological image data and yields reconstructions with higher perceptual image quality than previous methods. We show that for the evaluation of image quality in the WSI domain, several aspects are important and demonstrate that none of the existing metrics is ideally suited to compare compression schemes, but the combination of MSE, MS-SSIM and LPIPS can be a helpful. Future work could focus on further specialized metrics for WSI, which could evaluate morphological characteristics of cells as a quality metric. Also, generative decoding methods, such as GAN- or diffusion-based decoders \cite{JensPetersen,DiffusinRecent1,DiffusionRecent2}, show promising results in compensating for information loss on the encoder side. Nevertheless, practical considerations for both the encoder and decoder must be taken into account. In clinical settings, it is common to utilize powerful devices for image acquisition, while deploying subsequent lightweight image analysis components in the analysis pipeline. So far our method is ideally suited for HE stained images, which is the most common staining procedure in clinical practice, since the model from \cite{CIGA2022100198} is mostly trained on HE slides. Thus future work should also consider special models for deep supervision that are trained on special stains. 

\subsubsection{Acknowledgements} This work was partially supported by the DKTK Joint Funding UPGRADE, project “Subtyping of pancreatic cancer based on radiographic and pathological features“ (SUBPAN), and by the Deutsche Forschungsgemeinschaft (DFG, German
Research Foundation) under the grant 410981386.

%
%
%
%
\bibliographystyle{splncs04}
\bibliography{BibFile}

\begin{thebibliography}{10}
\providecommand{\url}[1]{\texttt{#1}}
\providecommand{\urlprefix}{URL }
\providecommand{\doi}[1]{https://doi.org/#1}

\bibitem{bmshj2018-factorized}
Ballé, J., Minnen, D., Singh, S., Hwang, S.J., Johnston, N.: Variational image compression with a scale hyperprior. In: International Conference on Learning Representations (2018), \url{https://openreview.net/forum?id=rkcQFMZRb}

\bibitem{begaint2020compressai}
B{\'e}gaint, J., Racap{\'e}, F., Feltman, S., Pushparaja, A.: Compressai: a pytorch library and evaluation platform for end-to-end compression research. arXiv preprint arXiv:2011.03029  (2020)

\bibitem{LungColonData}
Borkowski, A.A., Bui, M.M., Thomas, L.B., Wilson, C.P., DeLand, L.A., Mastorides, S.M.: Lung and colon cancer histopathological image dataset (lc25000) (2019). \doi{10.48550/ARXIV.1912.12142}, \url{https://www.kaggle.com/datasets/andrewmvd/lung-and-colon-cancer-histopathological-images}

\bibitem{chen_quantitative_2020}
Chen, Y., Janowczyk, A., Madabhushi, A.: Quantitative assessment of the effects of compression on deep learning in digital pathology image analysis. {JCO} clinical cancer informatics  \textbf{4},  221--233 (2020). \doi{10.1200/CCI.19.00068}

\bibitem{CIGA2022100198}
Ciga, O., Xu, T., Martel, A.L.: Self supervised contrastive learning for digital histopathology. Machine Learning with Applications  \textbf{7},  100198 (2022). \doi{https://doi.org/10.1016/j.mlwa.2021.100198}

\bibitem{Shannon}
Cover, T.M., Thomas, J.A.: Elements of information theory (2012)

\bibitem{ghazvinian_zanjani_impact_2019}
Ghazvinian~Zanjani, F., Zinger, S., Piepers, B., Mahmoudpour, S., Schelkens, P.: Impact of {JPEG} 2000 compression on deep convolutional neural networks for metastatic cancer detection in histopathological images. Journal of Medical Imaging  \textbf{6}(2), ~1 (2019). \doi{10.1117/1.JMI.6.2.027501}

\bibitem{JensPetersen}
Ghouse, N.F.K.M., Petersen, J., Wiggers, A.J., Xu, T., Sautiere, G.: Neural image compression with a diffusion-based decoder (2023), \url{https://openreview.net/forum?id=4Jq0XWCZQel}

\bibitem{KatherDataSet}
{Kather}, J.N., {Weis}, C.A., {Bianconi}, F., {Melchers}, S.M., {Schad}, L.R., {Gaiser}, T., {Marx}, A., {Z{\"o}llner}, F.G.: {Multi-class texture analysis in colorectal cancer histology}. Scientific Reports  \textbf{6},  27988 (Jun 2016). \doi{10.1038/srep27988}, \url{https://www.kaggle.com/datasets/kmader/colorectal-histology-mnist}

\bibitem{Hific}
Mentzer, F., Toderici, G.D., Tschannen, M., Agustsson, E.: High-fidelity generative image compression. In: Larochelle, H., Ranzato, M., Hadsell, R., Balcan, M., Lin, H. (eds.) Advances in Neural Information Processing Systems. vol.~33, pp. 11913--11924. Curran Associates, Inc. (2020)

\bibitem{DiffusinRecent1}
Pan, Z., Zhou, X., Tian, H.: Extreme generative image compression by learning text embedding from diffusion models. ArXiv  \textbf{abs/2211.07793} (2022)

\bibitem{Multiplebpp1}
Rippel, O., Anderson, A.G., Tatwawadi, K., Nair, S., Lytle, C., Bourdev, L.D.: Elf-vc: Efficient learned flexible-rate video coding. 2021 IEEE/CVF International Conference on Computer Vision (ICCV) pp. 14459--14468 (2021)

\bibitem{russakovsky_imagenet_2015}
Russakovsky, O., Deng, J., Su, H., et~al.: {ImageNet} large scale visual recognition challenge  (2015), \url{http://arxiv.org/abs/1409.0575}

\bibitem{Multiplebpp2}
Song, M.S., Choi, J., Han, B.: Variable-rate deep image compression through spatially-adaptive feature transform. 2021 IEEE/CVF International Conference on Computer Vision (ICCV) pp. 2360--2369 (2021)

\bibitem{BreaKHis}
Spanhol, F.A., Oliveira, L.S., Petitjean, C., Heutte, L.: A dataset for breast cancer histopathological image classification. IEEE Transactions on Biomedical Engineering  \textbf{63}(7),  1455--1462 (2016). \doi{10.1109/TBME.2015.2496264}

\bibitem{TheisPioneering}
Theis, L., Shi, W., Cunningham, A., Husz{\'a}r, F.: Lossy image compression with compressive autoencoders. In: International Conference on Learning Representations (2017), \url{https://openreview.net/forum?id=rJiNwv9gg}

\bibitem{MSSSIM}
Wang, Z., Simoncelli, E., Bovik, A.: Multiscale structural similarity for image quality assessment. vol.~2, pp. 1398 -- 1402 Vol.2 (12 2003). \doi{10.1109/ACSSC.2003.1292216}

\bibitem{Multiplebpp3}
Wu, L., Huang, K., Shen, H.: A gan-based tunable image compression system. 2020 IEEE Winter Conference on Applications of Computer Vision (WACV) pp. 2323--2331 (2020)

\bibitem{Vimeo}
Xue, T., Chen, B., Wu, J., Wei, D., Freeman, W.T.: Video enhancement with task-oriented flow. International Journal of Computer Vision (IJCV)  \textbf{127}(8),  1106--1125 (2019)

\bibitem{DiffusionRecent2}
Yang, R., Mandt, S.: Lossy image compression with conditional diffusion models. arXiv preprint arXiv:2209.06950  (2022)

\bibitem{LPIPS}
Zhang, R., Isola, P., Efros, A.A., Shechtman, E., Wang, O.: The unreasonable effectiveness of deep features as a perceptual metric. In: Proceedings of the IEEE Conference on Computer Vision and Pattern Recognition (CVPR) (June 2018)

\end{thebibliography}

\end{document}